\documentclass[twocolumn,showpacs,preprintnumbers,amsmath,amssymb]{revtex4}

\usepackage{graphicx}
\usepackage{dcolumn}
\usepackage{bm}
\usepackage{amssymb}
\usepackage{enumerate}

\newcommand\kslinv{K_S,K_L \to invisible}
\newcommand\kslbinv{K_S (K_L) \to invisible}
\newcommand\ksl{K_{S,L} \to invisible}
\newcommand\ksinv{K_S \to invisible}
\newcommand\klinv{K_L \to invisible}
\newcommand\koinv{K^0 \to invisible}

\newcommand\kob{\overline{K}^0}

\def\address{\@ifstar{\address@star}%
  {\@ifnextchar[{\address@optarg}{\address@noptarg}}}

\begin{document}

\author{S.N.~Gninenko}
\affiliation{Institute for Nuclear Research, Moscow 117312, Russia}


\title{Search for invisible decays of $\pi^0, \eta, \eta', K_S$ and $K_L$: A probe of new physics and \\
tests using the Bell-Steinberger relation } 

\date{\today}

\begin{abstract}
In the standard model the rate of the $\pi^0, \eta, \eta', K_S, K_L\to \nu \overline{\nu}$ decays is predicted to be   extremely small. Therefore, an observation of any of these  mesons ($M^0$) decaying into an invisible final state  would unambiguously signal the presence  of new physics.      
The Bell-Steinberger relation connects CP and CPT violation in the  mass matrix to CP and CPT violation in all decay channels of neutral kaons. It is  a powerful tool for testing CPT invariance in the $K^0-\kob$ system,  
assuming that  there are no significant  undiscovered decay modes of either $K_S$ or $K_L$ which 
 could contribute to the precision of the results.
The $\kslinv$  decays have never been tested and the  question  of how
 much these  decays can influence the Bell-Steinberger analysis of the $K^0-\kob$ system still remains open.  In the present work we propose a new experiment to search for the  $M^0\to invisible$ decays which aims at  probing new physics and answering  this question. The experiment utilizes high-energy  hadronic  beams from the CERN SPS and  
the charge-exchange  reactions  of pions or kaons on nucleons of an active target, e.g. $\pi^- (K^-) + p\to M^0  + n $, as a source of  the well-tagged $M^0$s emitted in the forward direction with the beam energy.  
If the decay $M^0\to invisible$  exists, it could be observed   by looking for an excess of events with 
a specific  signature: the complete  disappearance of the beam energy in the detector. 
This unique signal of $M^0\to invisible$ decays allows for searches of the decays $\kslinv$
with a sensitivity in the branching ratio Br$(\kslbinv) \lesssim 10^{-8} (10^{-6})$, and $\pi^0,\eta, \eta' \to invisible$ decays with a sensitivity  a few  orders of magnitude beyond the present experimental limits.
 This experiment is complementary to the one   recently proposed for the search for
invisible decays of dark photons and fits well with  the present kaon physics program at CERN.    
\end{abstract}
\maketitle

\section{Introduction}
Experimental studies of   invisible decays, i.e. particle transitions to 
an experimentally unobservable final state, played an important role both in the development of the
standard model (SM) and in testing its extensions \cite{pdg}. 
It is worth remembering the precision measurements of the  $Z \to invisible$ decay rate for the determination of the number of lepton families in the SM.  
In recent years, experiments on invisible particle decays have  received considerable attention.
Motivated by  various models of physics beyond the SM, these experiments include searches for 
invisible decays of $\pi^0$ mesons at  E949  \cite{pi0}, $\eta$ and $\eta'$ mesons
 at BES \cite{bes14},
  heavy $B$-meson decays at Belle \cite{belle}, BaBAR \cite{babar}, and  BES \cite{bes},  and
 invisible decays of the upsilon(1S) resonance at CLEO \cite{cleo},  baryonic number violation with  nucleon disappearance at SNO \cite{sno}, BOREXINO \cite{borexino}, and  KamLAND \cite{kamland}, see also Ref.\cite{tretyak}, electric-charge-nonconserving electron decays $e^- \to invisible$  \cite{klap},   neutron-mirror-neutron oscillations at PSI \cite{psinn} and  the ILL reactor \cite{ser}, and the disappearance of neutrons into another brane world \cite{sar}. 
One could also mention experiments looking for extra dimensions with invisible decays of positronium \cite{gkr,bader}, and proposals for new   experiments to  search for muonium annihilation into two neutrinos,
 $\mu^+ e^- \to \nu \overline{\nu}$ \cite{muonium}, electric charge nonconservation in the muon decay  
 $\mu^+ \to invisible$  \cite{sngmu},  and mirror-type dark matter through the 
invisible decays of orthopositronium in vacuum \cite{paolo}.
 
The use of the (pseudo)scalar mesons ($M^0$), such as $\pi^0, \eta, \eta', K_S$, and  $K_L$, to search for 
new physics by looking for their decays into invisible final states is advantageous, because in the 
standard model the rate of the $\pi^0, \eta, \eta', K_S, K_L\to \nu \overline{\nu}$ decays is predicted to be   extremely small. For massless neutrino the decay $M^0 \to \nu \overline{\nu} $ is forbidden kinematically by angular momentum conservation. Indeed, in the $M^0$ rest frame the neutrinos produced in the decay fly away in  opposite directions along the same line. Since the neutrinos and antineutrinos are
massless, the projection of the sum of their spins on this line equals $\pm$1. The projections of the orbital angular momentum of the neutrino on this line are equal to
zero. Since in the initial state we have a scalar, the process is forbidden. For the case of massive neutrinos
 their spins in the rest frame  must be opposite and, hence, one of the them is forced to  have the  "wrong" helicity. This results in  the $M^0\to \nu \overline{\nu}$ decay rate being proportional to the neutrino mass squared: 
\begin{equation}
\Gamma (M^0\to \nu \overline{\nu}) \sim \Bigl(\frac{m_\nu}{m_{M^0}}\Bigr)^2 \lesssim 10^{-16} 
\end{equation}
Thus, we see that, if the decay $M^0\to invisible$ is observed it would unambiguously signal the presence  of new physics, which could be due to, e.g. the existence of a new gauge boson with nonuniversal couplings to quarks, or the $M^0$ transitions into a hidden sector, or other effects \footnote{S.N. Gninenko, N.V. Krasnikov. Paper in preparation}. 

Another motivation for  searching for (in particular) the $\kslinv$ decays is related to additional tests
of the $K^0-\overline{K}^0$ system using the Bell-Steinberger relation\cite{bs}. 
This relation, obtained by using the unitarity condition, connects
 CP and CPT violation in the mass matrix of the kaon system, i.e.  parameters describing T and CPT noninvariance,  to CP and CPT violation in all decay channels of neutral kaons,  see, e.g. Refs. \cite{js, adpdg, lm, dafne, bloch}. 
We know that only CPT appears to be an exact symmetry of nature, while C, P and T are known to be violated. Hence, testing the validity of CPT invariance  probes the basis of the standard model. The Bell-Steinberger relation remains the most sensitive 
  test of CPT symmetry. For example,  the  analyses of the KLOE Collaboration have reached the impressive sensitivity of $-5.3\times  10^{-19}$ GeV $<m_{K^0} -  m_{\overline{K}^0}< 6.3\times 10^{-19}$ GeV at 95\% C.L. for the neural kaon mass difference \cite{kloe}; see also Ref. \cite{cplear}.

Briefly,  within the Wigner-Weisskopf approximation, the time evolution of the neutral kaon system is described by \cite{kloe}:
\begin{equation}
i\frac{d\Phi(t)}{dt}=H\Phi(t)=\Bigl(M-i\frac{i}{c}\Gamma\Bigr) \Phi(t)
\label{kk}
\end{equation}
where $M$ and $\Gamma$ are $2\times 2$  Hermitian matrices, which are time independent,  and $\Phi(t)$ is a two-component state vector in the $K^0 - \overline{K}^0$ space. Denoting by $m_{ij}$ and $\Gamma_{ij}$ the elements of $M$ and $\Gamma$ in the  $K^0 - \overline{K}^0$basis, $CPT$ invariance implies
\begin{eqnarray}
m_{11}=m_{22}~~ (\rm{or} ~m_{K^0}=m_{\overline{K}^0})~ \rm{and} \\ \nonumber
 \Gamma_{11}=\Gamma_{22}~~  (\rm{or} ~\Gamma_{K^0}=\Gamma_{\overline{K}^0})
\end{eqnarray}

The eigenstates of Eq. (\ref{kk}) can be written as
\begin{eqnarray}
K_{S,L}= \frac{1}{\sqrt{2(1+|\epsilon_{S,L}|^2)}}\Bigl((1+\epsilon_{S,L})K^0  \nonumber \\
\pm (1-\epsilon_{S,L})\overline{K}^0) \\ \nonumber
\end{eqnarray}
with
\begin{eqnarray}
\epsilon_{S,L}  = \frac{1}{m_L-m_S+i(\Gamma_S -\Gamma_L)/2}\Bigl[-i\rm{Im}(m_{12})-\\ \nonumber
\frac{1}{2}\rm{Im}(\Gamma_{12}) \pm \frac{1}{2}(m_{\overline{K}^0}-m_{K^0}-\frac{i}{2}(\Gamma_{\overline{K}^0}-\Gamma_{K^0})\Bigr] \equiv \epsilon \pm \delta 
\end{eqnarray}
The unitarity condition allows us  to express the four elements of $\Gamma$ in terms of appropriate combinations of the 
kaon decay amplitudes $A_i$:
\begin{equation}
\Gamma_{ij}=\sum_f A_i(f)A_j^*(f), ~~i,j=1,2=K^0,\kob
\end{equation}
where the sum is over all the accessible final states. 
\begin{eqnarray}
\Bigl(\frac{\Gamma_S+\Gamma_L}{\Gamma_S-\Gamma_L}+i\rm{tan}\phi_{SW}\Bigr) \Bigl(\frac{Re(\epsilon)}{1+|\epsilon|^2}-i\rm{Im}(\delta)\Bigr)  \nonumber \\
=\frac{1}{\Gamma_S-\Gamma_L}\sum_F A_L(f)A_S^*(f),
\label{bsr}
\end{eqnarray}
where $\phi_{SW} = \rm{arctan}[2(m_L-m_S)/(\Gamma_S-\Gamma_L)]$.  One can see that the Bell-Steinberger relation \eqref{bsr} relates a possible violation of CPT invariance ($m_{K^0}= m_{\overline{K}^0}$ and/or 
$\Gamma_{K^0}=\Gamma_{\overline{K}^0}$) in the $K^0-\overline{K}^0$ system to the observable CP-violating interference of $K_S$ and $K_L$ decays into the same final state $f$. If CPT invariance is not violated, 
then  $Im(\delta)=0$. We stress that  any evidence for $Im(\delta)\neq 0$ resulting from this relation can only manifest the violation of CPT or unitarity \cite{adpdg}. 

Generally, the advantage of the neutral kaon system is attributed to the fact, that only a few (hadronic)  decay modes give significant contributions to the rhs of Eq. (\ref{bsr}). In particular, it is assumed  
 that there are no significant contributions from invisible  decay modes of either 
$K_L$ or $K_S$ which, however, have never been   experimentally tested.
Therefore,  the contribution from these decay modes and  how much the errors on 
$Re(\epsilon)$ and $Im(\delta)$ would  increase if the invisible modes have maximal CP violation
are  still open questions, see, e.g.,  Ref.\cite{worksh}. As long as these questions are not answered experimentally, further  tests of CPT symmetry via Bell-Steinberger relations remain important.

In order to estimate the contribution from the $\kslinv$ decay to the right-hand side of Eq.(\ref{bsr}) one can 
follow the same procedure used by the KLOE Collaboration for the estimation of the contribution from the 
$K_L,K_S \to \pi^0 \pi^0 \pi^0 $ channel to  Eq.(\ref{bsr}) \cite{kloe}.  For invisible states we define  
\begin{eqnarray}
\alpha_{inv}\equiv \frac{1}{\Gamma_S}<A_L(inv)A^*_S(inv)> =  \\ \nonumber
\frac{\tau_{K_S}}{\tau_{K_L}} \eta^*_{inv} Br(K_L\to invisible)
\end{eqnarray}
where  $\eta_{inv}$ coefficient is the $A_L(inv),A_S(inv)$ amplitude ratio.
As there is no  experimental limit on $\eta^*_{inv}$ it would be more convenient to set a bound on 
$|\alpha_{inv}|$ by using the relation 
\begin{equation}
|\alpha_{inv}|^2 = \frac{\tau_{K_S}}{\tau_{K_L}} Br(K_L\to invisible)Br(K_S\to invisible)
\label{bsinv}
\end{equation}
and  experimental constraints on the rate of $\kslinv$ decays (derived below). Equation (\ref{bsinv}) is based on the assumption 
that the $K_L (K_S)  \to invisible$ decay mode is dominated by a single  $CP$-conserving (-violating) amplitude.
Note that all quantum numbers of the invisible final state
have to be equal between $K_L$ and $K_S$ decays in order to allow for interference between two amplitudes on the 
right-hand side of Eq.(\ref{bsr}). 

If the decays $\kslinv$ exist they will contribute to the total $K_L, K_S$ decay rate
\begin{eqnarray}
\Gamma_{K_L (K_S)}= \sum_i \Gamma_i(K_L(K_S) \to visible) \\ \nonumber
 + \Gamma(K_L(K_S) \to invisible)
 \label{tot}
\end{eqnarray}
resulting in 
\begin{equation}
\sum_i Br_i(K_L(K_S) \to visible) < 1 
\end{equation}
In order to  obtain bounds on the branching fraction $Br(K_L (K_S) \to invisible)$, to estimate the allowed extra contribution of $\kslinv$ decays to Eq. (10), and to derive a limit on $|\alpha_{inv}|^2$, we use the results of the most precise  measurements of the branching fractions of the visible 
$K_S,~ K_L$ decay modes from Particle Data Group PDG \cite{pdg}. Summing up all measured branching fractions,  we get 
\begin{equation}
\sum_i Br_i(K_S \to visible) = 1.00068 \pm 0.00048 
\end{equation}
and 
\begin{equation}
\sum_i Br_i(K_L \to visible) = 1.00032 \pm 0.00058 
\end{equation}
 resulting, respectively,   in 
 \begin{equation}
 Br(K_S \to invisible) < 1.1 \times 10^{-4}, ~ ({\rm  ~95\%~ C.L.}),
  \label{brks}
 \end{equation}
  and 
 \begin{equation}
 Br(K_L \to invisible) < 6.3 \times 10^{-4}, ~({\rm ~95\%~ C.L.}).
 \label{brkl}
 \end{equation}

Using for the $K_S$ and  $K_L$ lifetime the averages $\tau_{K_S}=0.08958\pm 0.00006$ ns  and $\tau_{K_L}= 51.16\pm 0.2$ ns, we obtain 
\begin{equation}
|\alpha_{inv}| < 2.8 \times 10^{-5}, ~ ({\rm  ~95\%~ C.L.}).
\label{bsinvlim}
\end{equation} 
Note that the averages and errors for the branching fractions for the visible $K_S,~ K_L$ decay modes
were obtained by PDG with a fit procedure. For comparison, we also try to estimate $|\alpha_{inv}|$ 
by using results  from direct measurements of $Br_i(K_S (K_L) \to visible)$, obtained mostly by the KLOE and NuTeV Collaborations \cite{pdg}. For the $K_S$ and  $K_L$ lifetimes
quoted above the new limit on $|\alpha_{inv}|$ is similar to the one from Eq.(\ref{bsinvlim}) within a factor of 2.  

It is interesting to compare these  results  with the limit on the contribution of the $3\pi^0$ decay modes to the Bell-Steinberger relation derived by the KLOE Collaboration, $|\alpha_{\pi^0 \pi^0 \pi^0 }| < 7 \times 10^{-6}$ at 95\% C.L. \cite{kloe}. It should be noted,  that $K_S,K_L$ decay parameters  from the  decay channels with the branching ratio Br$(K_S \to  f)= \Gamma(K_S\to f)/ \Gamma_S \gtrsim 10^{-5}$ and   Br$(K_L \to  f)\Gamma_L/ \Gamma_S  \gtrsim 10^{-5}$  are 
 within the present accuracy of Eq.(\ref{bsr}) and contribute to the Bell-Steinberger analysis of the kaon system \cite{adpdg}. Therefore, an improvement of the  bounds of Eq.(\ref{brks} and \ref{brkl}) by at least an order of magnitude is necessary  in order to agree with the present accuracy  of Eq.(\ref{bsr}).
  The decay mode that gives the largest contribution to the precision of the Bell-Steinberger analysis at the 
 level $O(10^{-5})$  is now $K_L \to \pi^+ \pi^-$ through the uncertainty on the phase $\phi +-$ \cite{kloe}. 

One of the aims of this work  is to show that the limits of Eqs.(\ref{brks},\ref{brkl}) can be significantly improved by the proposed experiment that would search for the still unexplored decay modes $\kslinv$
  with a high-energy $K^\pm$ beam at the CERN SPS. The expected sensitivity 
in the branching fraction is  Br$(\kslbinv) \lesssim 10^{-8} (10^{-6})$. 
The experiment is  also capable of  a sensitive search for $\pi^0, \eta, \eta' \to invisible $ decays
and could improve the existing limits by more than an order of magnitude. 
  The rest of the paper is organized as follows. The method of the search and the experimental setup are 
  described  in Sec. II, the background sources are discussed in Sec. III, and   the expected sensitivity  for the decay $M\to invisible$ is presented in Sec. IV. Section V contains concluding remarks. 
 
\begin{figure*}[tbh!]
\includegraphics[width=.95\textwidth]{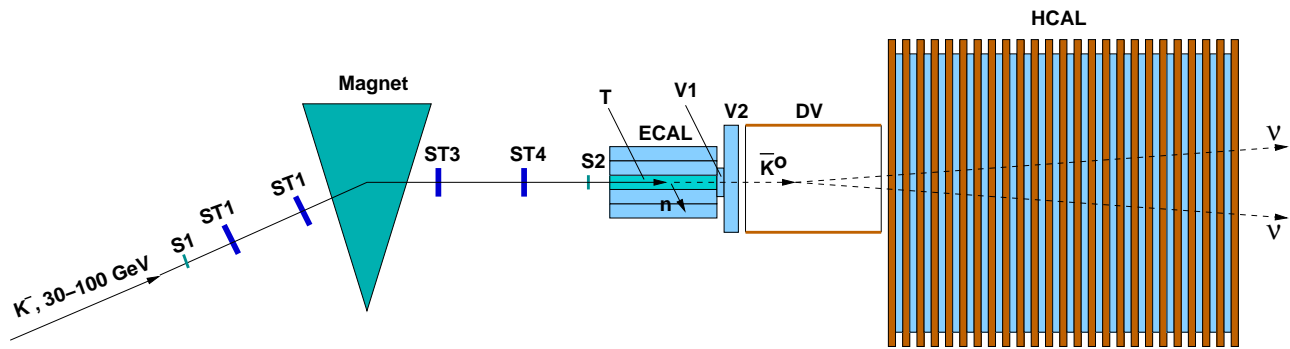}
\caption{\label{fig:setup} 
Schematic illustration of the setup to search for the invisible decays of neutral 
kaons in the proposed  experiment at high energies. The beam of incident charged kaons is defined by the scintillating counters S$_{1,2}$. The momentum of the beam is additionally selected  with a momentum spectrometer
consisting of a dipole magnet and a low-density tracker, made of a set of  straw tube chambers (ST1-ST3) or Micromegas detectors.   
The $K^0$s are produced in the charge-exchange reaction of kaons scattering off nuclei 
in the active target T. The T is  surrounded by the Veto system consisting of the electromagnetic calorimeter (ECAL) and two high-efficiency scintillating counters V1 and V2 used against photons or charged secondaries that could escape the target at a large angle or in the forward direction, and a massive completely hermetic hadronic calorimeter (HCAL) 
to absorb the energy of all secondaries. 
The $K^0$s either  decay invisibly  in the target, or they  penetrates  V1 and  V2     without interactions and (as shown)   decay invisibly in flight in the decay volume DV into, e.g., a pair of heavy $\nu$, which carry away almost  all of the primary beam energy, resulting in the zero-energy   signature in the detector.
The recoil neutron shown typically deposits  a small amount of energy. The same setup can be used for the searches 
for the $\pi^0, \eta, \eta' \to invisible $ decays (see text. }
\label{setup}
\end{figure*}   

\section{An experiment to search for  $\pi^0, \eta, \eta', \koinv$ decays} \label{sec:ExpInvisible}

The  detector specifically designed to search for the  $\pi^0,\eta,\eta',\kslinv$ decays is schematically shown in 
Fig. \ref{setup}. This experimental setup  is complementary to the one   recently proposed for the search for
invisible decays of dark photons at the SPS at CERN \cite{sngldms, ldms}.    
The experiment could employ, e.g., the  H4 hadron beam, which is produced in the target T2 of the SPS and transported to the detector in an evacuated beamline tuned to a freely adjustable  beam momentum from 10-300 GeV/c \cite{sps}. 
The typical maximal beam  intensity at $\simeq$ 50-100 GeV, is of the order of $\simeq 10^7~\pi^\pm$ 
and $\simeq 10^6~ K^\pm$ for one  SPS spill with $10^{12}$ protons on target. The typical SPS cycle for 
fixed-target (FT) operation lasts 14.8 s, including 4.8 s  spill duration. The maximal number of FT cycles is four per minute.  The 
beam has high purity: the   admixture of the other charged particles is below $10^{-2}$. The beam can be focused onto a spot of the order of a few cm$^2$. 

We first consider the experiment that would search for the $\kslinv$ decays. 
The method of the search is as follows. The source of $K^0(\overline{K}^0)$s is the charge-exchange reaction of high-energy kaons on nucleons of an  active target
\begin{eqnarray}
 K^- + p \to \overline{K}^0 + n,& ~\text{or}  \nonumber \\
K^+ +n \to K^0 +p&     
\label{kchex}
\end{eqnarray}
where the neutral kaon is emitted mainly in the forward direction with the beam momentum and the 
recoil nucleon carries away a small fraction of the beam energy.
Further, we will assume  no difference between these two reactions. 
The  invisible decay $\koinv$ is expected to be a very rare event which  occurs with a rate much smaller  
than the total $K^0$ production rate. Hence, its observation presents a challenge for the design and performance
of the detector. 
 
 The detector shown in Fig. \ref{setup} is equipped with the   scintillating counters S1 and  S2 (which define the beam),  an active target $T$  surrounded by a  high-efficiency  electromagnetic calorimeter (ECAL) serving  as a veto 
 against photons and other secondaries emitted from the target  at large angles, high-efficiency forward veto counters V1 and V2, a decay volume DV, and a massive, hermetic hadronic calorimeter (HCAL) located at the end of the setup to detect energy deposited by secondaries from the  primary  interactions $K^\pm A \to$ anything 
 of $K^\pm$s with nuclei $A$  in the target. For searches at low  energies,  Cherenkov counters  to enhance the incoming hadron tagging efficiency can be used.

 The reaction~\eqref{kchex} occurs practically uniformly over the length of the target. 
 The  fraction of the primary kaon (pion) energy deposited in the target is used to determine the position
of the interaction vertex along the beam direction.
 The produced
   $K^0$ - composed of equal portions of $K_S$ and $K_L$-  either 
decay quickly in the target $T$, or  penetrates the veto system without interactions  and either decays in flight  in the DV  or interacts in the HCAL. If the $K_S$ and $K_L$ decay invisibly, it is 
assumed that  the final-state  particles in this case also penetrate the rest of the detector without prompt decay  
into ordinary particles, which could deposit energy in the HCAL. 
  In order to suppress  background due to the detection inefficiency,  the detector must be longitudinally completely hermetic. To enhance detector hermeticity, the hadronic calorimeter has a total thickness of 
   $\simeq 28 ~\lambda_{int}$ (nuclear interaction lengths). 

The occurrence of $\kslinv$ decays produced in $K^\pm$ interactions would appear as an excess of events with a signal in the $T$, see Fig.~\ref{setup} and zero energy deposition in the rest of the detector (i.e. above that expected from the background sources). Thus, the signal candidate events have the signature 
\begin{equation}
S_{\koinv} = {\rm T \cdot \overline{V1\cdot V2\cdot HCAL}}
\label{signinv} 
\end{equation}
and should satisfy the following selection criteria. 
\begin{enumerate}[(i)]
\item The measured momentum of the incoming kaon should correspond to its selected value.
\item The kaon should enter the target and the  interaction vertex  should be localized  within the target volume.
\item The should be no energy deposition in  the ECAL veto, V1 and V2.   
\item The fraction of the beam energy deposited in the HCAL modules should be consistent with zero.
\end{enumerate}

The application of all the previous   considerations to the search for the decays $\pi^0,\eta, \eta' \to invisible$ with the 
same detector  is straightforward. The source of $\pi^0,\eta, \eta'$ is the charge-exchange reaction of high energy pions on nucleons of the target
\begin{eqnarray}
 \pi^- + p \to \pi^0, \eta, \eta' + n,&~ \text{or}  \nonumber \\    
  \pi^+ + n \to \pi^0, \eta, \eta' + p,&
\label{pichex}
\end{eqnarray}
where the final-state neutral meson  is emitted mainly in the forward direction with the beam momentum and the 
recoil nucleon carries away a small fraction of the beam energy. Similar to the $K$-meson case,  
 the occurrence of $\pi^0, \eta, \eta' \to invisible $ decays produced in $\pi^\pm$ interactions would appear as an excess of events with the same signature of 
 Eq.({\ref{signinv}), i.e. the presence of a signal in the $T$, see Fig.~\ref{setup} and zero energy deposition in the rest of the detector (i.e. above that expected from the background sources).

In Fig.\ref{pedmu} the expected distribution of the signal of Eq.(\ref{signinv}) in the HCAL is shown, i.e. the pedestal sum over the HCAL modules, representing the signal from an invisible final state  in units of hadronic energy. The FWHM of the signal peak is expected to be $\lesssim 100$ MeV. The distribution was estimated from the real data taken 
at the H4 beam. The simulated distribution of the energy deposited  in the  HCAL  by traversing muons with energy $E_{\mu}=100$ GeV is also shown for comparison. 
  
\section{Background}
The background reactions resulting in  the signature of the process $\pi,K + p \to M^0 + n;~ M^0 \to invisible$, see
 Eq. (\ref{signinv}),  can be classified as being due to physical-  and  beam-related sources.  To perform a full detector simulation in order  to investigate these backgrounds down to the level  $ \lesssim 10^{-10}$  would require  a prohibitively large amount of computer time. Consequently, only the following sources of background - identified as the most dangerous - are considered and evaluated  with  reasonable statistics combined  with numerical calculations:
\begin{figure}
\includegraphics[width=0.5\textwidth]{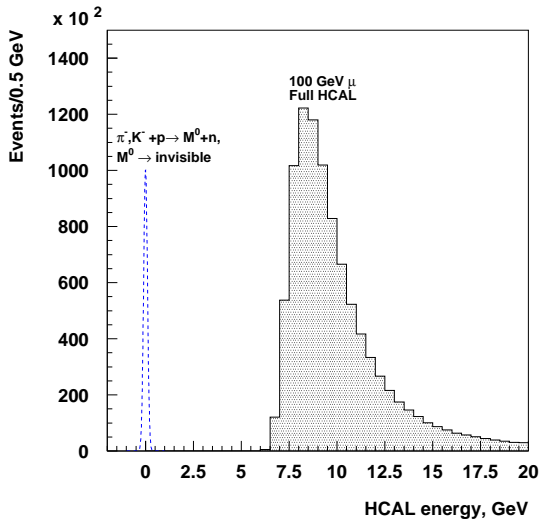}
\caption{The distribution of the energy deposited  in the  HCAL  by traversing muons with energy $E_{\mu}=100$ GeV. The peak of the pedestal sum over the HCAL is also shown,  representing the signal from the invisible final state  in units of hadronic energy. The FWHM of the signal peak is expected to be $\lesssim 100$ MeV.}
\label{pedmu}
\end{figure}

\begin{enumerate}[(i)]
\item One of the main background sources is related to the  low-energy tail in the distribution of the energy of the primary  hadronic beam. This tail is caused by the beam interactions with a passive material, such as the entrance windows of the beam lines, residual gas, etc. Another source of low-energy hadrons  is due to their decays in flight in the beam line when the low energy decay pions or muons mimic the signature Eq. (\ref{signinv}) in the detector.  
The uncertainties arising from the lack of knowledge of the dead material composition in the beam line  are potentially the largest source of systematic uncertainty in accurate calculations  of the fraction  and energy distribution of these events. An estimation shows  that the fraction of events with energy below  $\lesssim 10$ GeV in the hadron beam tuned, e.g.,  to 50 GeV could be as large as  $10^{-8}-10^{-6}$. Hence, the sensitivity of the experiment could be  determined by the presence of such particles
 in the beam, unless one takes special  measures to suppress this background.
To improve the primary high-energy hadron selection  and suppress background from the possible admixture of 
low-energy particles, one can use a tagging system utilizing  the magnetic spectrometer installed upstream of the detector,  as schematically shown in Fig.~\ref{setup}. 
\begin{figure*}
\includegraphics[width=0.5\textwidth]{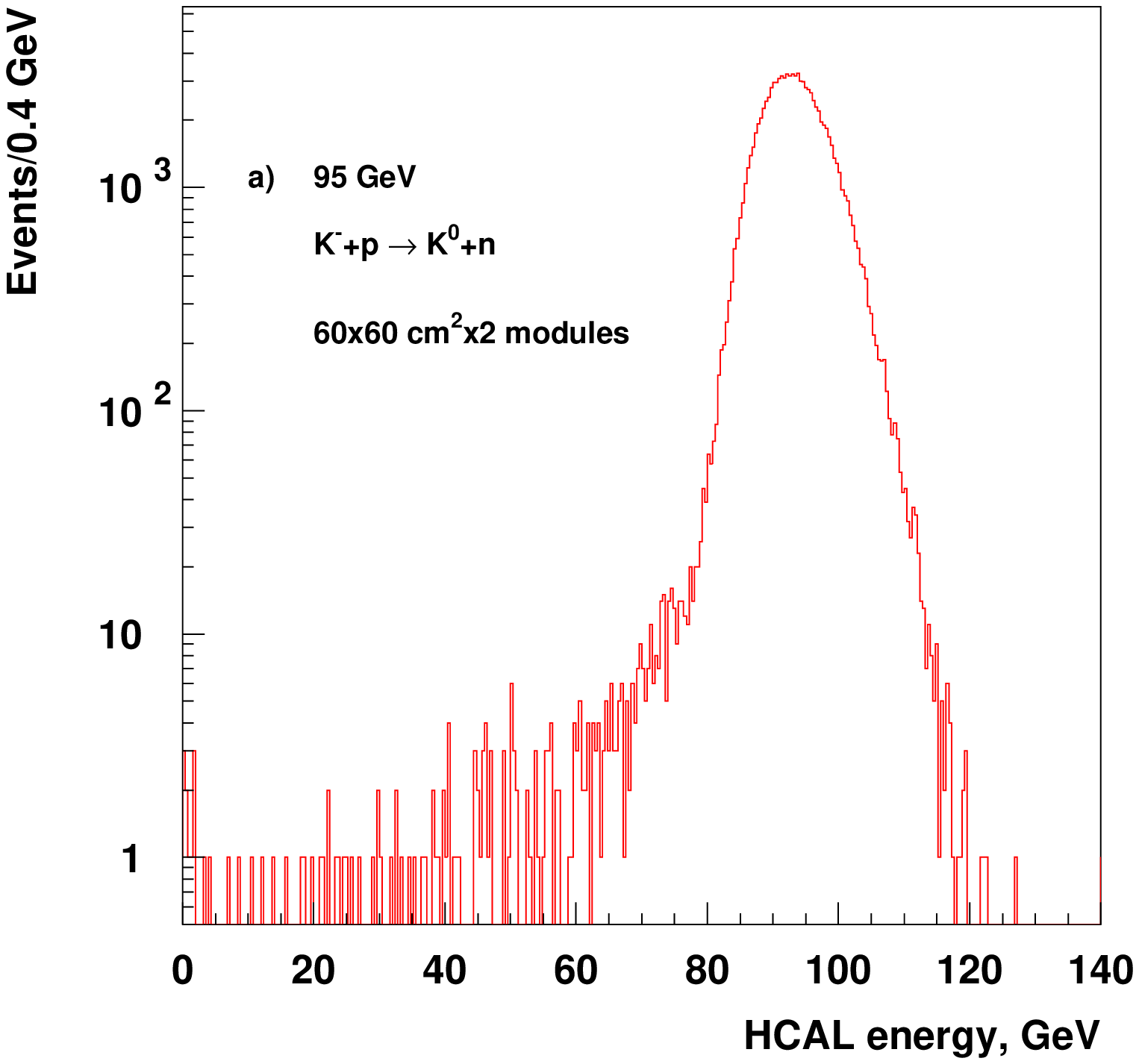}
\includegraphics[width=0.5\textwidth]{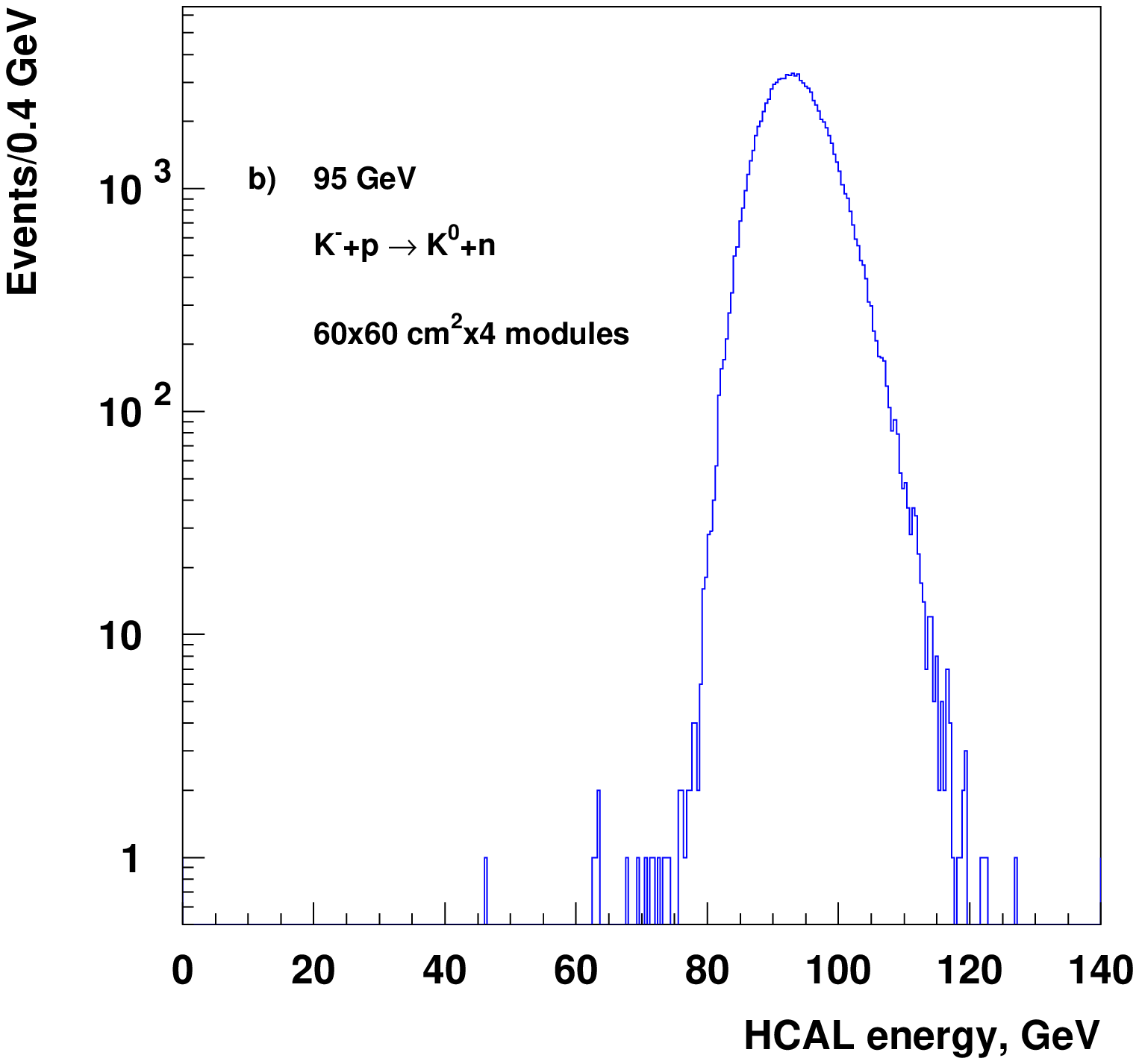}
\caption{  Expected distributions of energy deposited by $K^0$s with energy $\simeq$ 95 GeV from the charge exchange reaction Eq.(\ref{kchex})  in two (a)  and four (b) consecutive HCAL modules. The peak at zero energy in spectrum (a) is due to the punch-through neutral kaons.
}
\label{punch}
\end{figure*}

\item
The fake signature of Eq.\eqref{signinv} could also arise when the $K^0$ from the reaction \eqref{kchex} or a leading hadron, $h$,  from the reaction $\pi+A \to h + ... $ that occurred in the target is not detected due to the incomplete hermeticity of the HCAL. In this case, the  produced  $K^0$  punches through the HCAL without depositing energy above a certain threshold $E_{th}$. This effect is illustrated in  Fig. \ref{punch}(a), which shows the distribution 
of energy deposited by $K^0$s produced at 95 GeV, in two consecutive HCAL modules ($\simeq 14\lambda_{int}$). The distribution  is  obtained with GEANT4 simulations \cite{geant}.  The peak of events at zero energy in the spectrum  is caused by  the punch-through neutral kaons. Those  events with a sum of energy released in two  HCAL modules below the threshold 
$E_{th}\simeq 0.1$ GeV are considered as zero-energy events. In Fig. \ref{punch}(b),  one can also see that a similar distribution of  energy deposited by $K^0$s in four consecutive HCAL modules ($\simeq 28\lambda_{int}$) has no such zero-energy events.

The punch-through  probability is defined roughly by $\simeq exp(-L_{HCAL}/\lambda_{int})$,
where $L_{HCAL}$ is the HCAL thickness. It is $ \simeq 10^{-12}$ for the total thickness of the HCAL  about 
28 $\lambda_{int}$. Since performing detector simulations at this level of precision is not possible, the  rough estimate of the HCAL nonhermeticity for high-energy hadrons was cross-checked  with  GEANT4-based simulations in the following way. The low-energy tail in the  distribution of energy deposited in the full HCAL by 
$\simeq 10^7$ simulated neutral kaons  was fitted by a smooth polynomial function and extrapolated to the 
low-energy region  in order to evaluate the number of events below a certain threshold $E_{th}$, see Ref. \cite{ldms} for more
details. This procedure results in an estimate of the 
HCAL nonhermeticity, defined as the ratio of the number of events below the threshold $E_{th}$ to the total number of incoming particles, $(E<E_{th})/n_{tot}$. 
For example, for the energy threshold $E_{th} \simeq 0.5$ GeV the nonhermeticity is found  to be at the level $\simeq 0.4\times 10^{-11}$, which is in  satisfactory agreement with the above estimate when taking into account the accuracy of this procedure.  
This results in an overall conservative level for this background of $\lesssim 10^{-13}$ per incident beam kaon
reaction in the target. 

\item Another type of process which could mimic the tagging of the reaction 
$\pi,K + p \to M^0 + n;~ M^0 \to invisible$ and contribute to background is caused by in-flight $\pi,K\to \mu,e + \nu$  decays  of pions and kaons  after they have passed the spectrometer.
 The background of the low-energy  muon admixture  in the beam from the $\pi,K\to \mu \nu$  decays
  can be due to the following event chain. The decay muon entering the detector decays in flight into  a low-energy electron and a neutrino pair, $\mu \to e \nu \nu$ in the target. The electron then penetrates V1 and V2 without being detected, and deposits all its energy  in the HCAL, 
which is below the threshold $E_{th}\lesssim 0.5$ GeV. The probability for this  event chain  is found to be as small
as $P\lesssim 10^{-13}$. Similar background caused by the decays of the beam pions or kaons in the target was also found to be negligible. 

More dangerous are the in flight $\pi,K\to e + \nu$ decays   resulting in decay electrons with energy 
$\lesssim 1$ GeV, comparable with the energy deposited by the incoming $\pi,K$ in the target. To suppress this background a 2-3 $X_0$ preshower detector could be installed upstream  of the  target. The remaining part of this background is related to the in flight $\pi,K$ decays  in the target itself. The combined probability for such decay 
is suppressed by the small branching fraction of the decays Br$(\pi (K)\to e + \nu) < 10^{-4}(10^{-5})$ down to the 
level $\lesssim 10^{-9}(10^{-10})$.  Further suppression of this background could be achieved by 
using an active target, which is segmented along the beam axis with a separate readout of the signals. An additional suppression factor of 1 order of magnitude is expected form the analysis of the energy deposited in each segment, which should be comparable with the energy deposited by the minimum ionizing particle (MIP).

\item
The fake signature of Eq. (\ref{signinv}) could be due to the physical background: a muon scattering on a nucleon, e.g. 
  $\mu^- p \to \nu_\mu n$, accompanied by a poorly  detected neutron. Taking into account the corresponding cross section and the probability for the 
  recoil neutron to escape detection in the HCAL results in an overall level of this background 
  of $\lesssim 10^{-14}$ per incoming kaon.
\end{enumerate}

In Table~\ref{tab:table1} contributions from the all background processes are summarized for the primary $\pi^-$ and
$K^-$ beams with energy 40 and 95  GeV, respectively. The total background is found (conservatively) to be at the level $\lesssim 1.3 \times 10^{-12}$($\lesssim   10^{-12}$) per incoming  kaon (pion). 
Therefore, the search accumulated up to $\simeq 10^{12}$ $\pi^-$ or $K^-$ events is expected to be background free. The expected sensitivity in branching fractions is summarized below.  
\begin{table}[tbh!] 
\begin{center}
\caption{Expected contributions to the total level of background from different background sources estimated 
per incident $\pi^-$ and $K^-$ (see text for details).}\label{tab:table1}
\vspace{0.15cm}
\begin{tabular}{lr}
\hline
\hline
Source of background& Expected level\\
\hline
punch-through $K^0$s,    &$ \lesssim 10^{-13}$\\
 leading hadron from $\pi$ reactions&$ \lesssim 10^{-14}$\\
low-energy tail of the $\pi^-, K^-$ beam& $ \lesssim  10^{-12}$\\
HCAL nonhermeticity & $ \lesssim   10^{-13}$\\
$\pi^-,K^- \to \mu^- \nu$ decays in flight & $ \lesssim 10^{-13}$\\
$\pi^- (K^-) \to e^- \nu$ decays in flight & $ \lesssim 10^{-10}(10^{-11})$\\
$\mu^-$ induced  reactions & $\lesssim 10^{-14}$\\
\hline 
Total (conservative)  &         $ \lesssim  10^{-10}$ per $\pi^-$\\
  &         $ \lesssim    10^{-11}$ per $K^-$\\
\hline
\hline
\end{tabular}
\end{center}
\end{table}

\section{Expected sensitivity}

To estimate the sensitivity of the proposed experiment 
 a simplified feasibility study  based on GEANT4 \cite{geant}
Monte Carlo simulations have been  performed for 30-100 GeV pions and kaons. The ECAL is  the hodoscope  array of  the lead-scintillator  counters of the Shashlyk type counters ($X_0 \simeq 2$ cm) (see, e.g. Ref.\cite{shashlyk}),  each with the size of $36\times 36 \times 400$ mm$^3$, allowing for accurate measurements of the lateral energy leak from the target. The target is a block of radiation-hard plastic scintillator with  thickness $\simeq 0.5\lambda_{int}$ viewed by a photomultiplier.  
The veto counters are assumed to be  1-2 cm thick, high-sensitivity LYSO crystal arrays with a high light yield  of $\simeq 10^3$ photoelectrons per 1 MeV of deposited energy. It is also assumed that the veto's inefficiency  for the MIP  detection  is, conservatively, $\lesssim 10^{-4}$. The hadronic calorimeter is a set of four modules. 
Each module is a sandwich of alternating layers of iron and scintillator with thicknesses of 25 mm and 4 mm,  
respectively, and with a lateral size of $60\times 60$ cm$^2$.    Each module consists of 48 such layers and has 
a total thickness of $\simeq 7\lambda_{int}$. 
The number of photoelectrons produced by a MIP crossing the module  is in the range $\simeq$ 150-200 ph.e..
In Fig. \ref{pedmu} the  distribution of the energy deposited  in the  HCAL  by traversing muons with energy $E_{\mu}=100$ GeV is shown. The width of the lhs of the muon peak is defined by the fluctuations of the total 
number of collected photoelectrons $n_{ph.e.}\gtrsim 600$ ph.e. with a rms $\simeq 25$ ph.e.. This should be 
compared with the effective threshold $\simeq 100$ MeV, or $\simeq 8$ ph.e., for the zero-energy signal   
which is represented by the  distribution of the pedestal sum over the HCAL and also shown for comparison.
 Thus, one can see that the probability for an event with the MIP energy deposited in the HCAL to mimic the signal  due to fluctuations of $n_{ph.e.}$ is negligible. 
 
The hadronic energy resolution of the HCAL calorimeters as a function of the beam energy is taken to be 
$\frac{\sigma}{E} \simeq \frac{ 60 \%}{\sqrt{E}}$ \cite{ihephcal}. The energy threshold for the zero-energy in the HCAL is 0.1 GeV. The reported further analysis also  takes into account passive materials from  the DV vessel  walls.
\begin{figure}
\includegraphics[width=0.5\textwidth]{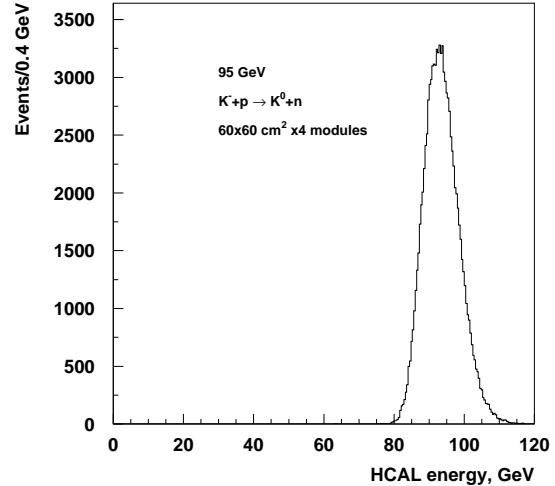}
\caption{ Expected distribution of the total  energy deposited by $K^0$s with energy $\simeq$ 95 GeV from the reaction \eqref{kchex})  in  four  HCAL modules.}
\label{koenergy}
\end{figure}
To estimate the expected sensitivities we used simulations of the process shown 
in Fig.\ref{setup} to calculate fluxes and energy distributions  of mesons produced in the target by taking into account  
the relative normalization of the yield of  different meson species $\pi^0 : \eta  : \eta'$ and $K^0$ from the original publications \cite{yud1, yud2}. 
The cross section of $\overline{K}^0$  production in the reaction \eqref{kchex} can be expressed as \cite{yud1}
\begin{eqnarray}
\frac{\sigma(K^-p\to \overline{K}^0 n)}{dt}\simeq (1-Gt)(exp[c_\rho t]+R^2exp[c_A t ]   \nonumber \\
-2R[cos\phi_+ -GT cos\phi_-]exp[(c_\rho + c_A)t/2]~~~~
\label{crsec}
\end{eqnarray}
where $t$ is the four-momentum transfer squared, $G= (33.5\pm 1.3)$ GeV$^{-2}$, $c_\rho=(15.5\pm0.3)$ GeV$^{-2}$, $c_A=(8.8\pm0.1)$ GeV$^{-2}$, $R=0.83\pm 0.05$, 
$cos\phi_+ = -0.08\pm 0.07$, and $cos\phi_+ = 0.23\pm 0.02$.
This formula   gives the parametric form of the charge-exchange cross sections  
for the production of neutral  kaons over the full phase space, up to $|t| \gtrsim 0.3$ GeV$^{-2}$.
For $\pi^0,\eta,\eta'$ we performed similar calculations by using the cross-section parametrization from 
Ref.\cite{yud2}. For the purpose  of this work,  the total $\pi^0, \eta, \eta'$ and $K^0$  production cross sections in 
the $\pi, K^-$ charge-exchange reactions in the target  were calculated  from thier linear extrapolation to the target atomic number. Note, that the yield of $\pi^0, \eta, \eta'$ and $K^0$ is also supposed to be measured 
{\it in situ} (see discussion below). Typically, the branching fractions of the charge-exchange reactions  are in the range
$\frac{\sigma(K^-p\to \overline{K}^0 n)}{\sigma(K^-p\to all)}\simeq \frac{\sigma(\pi^-p\to \pi^0 n)}{\sigma(\pi^-p\to all)}\simeq 10^{-4}-10^{-3}$ and depend on the beam energy \cite{yud1, yud2}.
\begin{figure}
\includegraphics[width=0.4\textwidth]{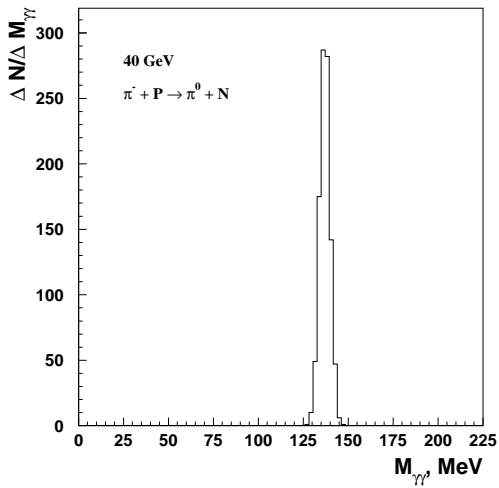}
\caption{Expected distribution of the diphoton invariant mass from the reaction \eqref{pichex}  in the hodoscopic ECAL, allowing accurate measurements of the photon coordinates, for the incoming pion energy of 40 GeV.  
The spectrum is peaked  at the $\pi^0$- mass of $\simeq$ 135 MeV. }
\label{pi0mass}
\end{figure}

The calculated fluxes and energy distributions  of mesons produced in the target are used to predict the number of signal events in the detector. 
For a given number of primary kaons $N_{K^-}$, the expected total number of 
$\ksl$ decays occurring within the decay length $L$ of the detector is given by 
\begin{equation}
n^{inv}_K = n^{inv}_{K_S} + n^{inv}_{K_L}
\label{ntot}
\end{equation}
with 
\begin{eqnarray}
&n^{inv}_{K_{S,L}}= k N_{K^-}Br(\ksl)  \nonumber \\
&\cdot \int\frac{\sigma(K^-p\to \overline{K}^0 n)}{dt} \Bigl[1-{\rm exp}\Bigl(-\frac{L M_{K^0}}{P_{K^0}\tau_{K_{S,L}}}\Bigr)\Bigr]\zeta \epsilon_{tag}  dt \nonumber \\
& \simeq  \zeta \epsilon_{tag} Br(\ksl) n^{dec}_{K_{S,L}}
\label{nev}
\end{eqnarray}
where coefficient $k$ is a normalization factor that was
tuned to obtain the total  cross section of the meson production,  $  P_{K^0}$ and $\tau_{K^0}$ are the  $K^0$ momentum and the lifetime of either $K_S$ or $K_L$ at rest, respectively, $\zeta$ is the signal reconstruction efficiency, $\epsilon_{tag}$ is the tagging efficiency of the final state,  and $n^{dec}_{K_{S,L}}$ is the total number of $K_{S,L}$ decays occurring  in the decay volume of length $L$. In this estimate we neglect the $K^0$ interactions in the target: the  average momentum of the incoming kaons is in the range $<p_{K^-}>\simeq 30-100$ GeV, the decay length   $L\simeq 10$ m, and the efficiency $\zeta \simeq 0.9$. The tagging efficiency $\epsilon_{tag}$ is  typically $\gtrsim 90\%$ \cite{yud1,yud2,gams}. The contributions from in flight $\pi,K \to \mu,e + \nu$ decays which could mimic the tagging is negligible. 
The inefficiency is caused mostly by the inelastic reactions, such as the inelastic charge-exchange reaction $\pi^- + p \to \pi^0 + N^{*0}$, with the subsequent
isobar decay $N^{*0}š\to n + \pi^0$, or  the process $\pi^- + p \to \pi^0 +\pi^0 + n$, or the reaction 
$\pi^- + p \to n'$s$+X$, etc...
All of these reactions are accompanied  in the final state by high-energy secondary particles emitted  in the forward direction    that are subsequently   absorbed  in the HCAL. Thus, the inefficient tagging of the neutral meson 
final state would not cause the missing energy background.      

In the case of no signal observation, the obtained results can be used to impose upper limits  on the previously discussed decays of $\pi^0, \eta, \eta', K^0$ into invisible final states; by using the relation $n^{inv}_K = n^{inv}_{K_S} + n^{inv}_{K_L}
< n^{inv}_{90\%} $, where $n^{inv}_{90\%}$ (= 2.3 events) is the 90$\%$ C.L. upper limit for the  number of signal events,  and Eq. (\ref{nev}), one can  determine 
the expected $90\%~ C.L.$ upper limits from the results of the proposed experiment. These bounds - calculated for 
the total number of $10^{12}$ incident pions or kaons and the background-free case - are summarized 
 in Table \ref{tab:table2}. The limits on Br($\ksinv$) and  Br($\klinv$) are obtained assuming 
 $n^{inv}_{K_L}=0$ and $n^{inv}_{K_S}$, respectively.
  Here we also assume that the exposure to the $\pi/K$ beam with the nominal rate is a few months,
 and  that the invisible final states do not decay promptly into the ordinary particles, which would deposit energy in the veto system or HCAL.

As mentioned before,  the yield of particles in the reactions  \eqref{kchex} and \eqref{pichex} used for normalization and limit calculations   is supposed to be determined in separate measurements with the same setup. In Fig. \ref{koenergy} an example of the expected distribution of energy deposited in the HCAL by neutral kaons from the reaction \eqref{kchex} is shown. The distribution is obtained with simplified simulations of $K^-$ interactions in the target and calculated for a beam  energy of 95 GeV and decay length  $L\simeq 0.1$ m. In this case, the decay length of $K_S$ and  $K_L$ is 
$ \gg L$ and kaons  mainly interact in the HCAL before they decay.   
  The candidate  events from the primary kaon interaction in the target are selected using the conditions of 
  Eq. (\ref{signinv}), but without the requirement of the absence of the energy deposition in the HCAL.
One can see that the $K^0$ energy spectrum is peaked at maximal beam energy.
  The distribution is almost background free, has a Gaussian shape and allows one to determine the $K^0$ yield 
with a good accuracy. The values of $n^{dec}_{K_S,L}$ entering  Eq.(\ref{nev}) can be easily determined by 
taking into account  that the $K^0$ is  composed of equal portions of $K_S$ and $K_L$ of known energies and lifetimes.   

The yield of $\pi^0, \eta$ and $\eta'$ mesons from the $\pi^-$ charge-exchange 
reaction in the target can also be determined using similar measurements. In this case, in order to 
improve the final-state identification,  a small hodoscope electromagnetic calorimeter that allows for the  
reconstruction of two photons from the $\pi^0, \eta, \eta' \to \gamma \gamma $ decays can be used. The yield of the 
$\pi^0, \eta$ and $\eta'$ mesons can be determined from the low background  peaks in the photon pair  mass spectrum corresponding  to the masses of the decay mesons. In  Fig.\ref{pi0mass} the simulated reconstructed distribution of diphoton invariant mass from the reaction $\pi^- +p \to \pi^0 + n$ in the  hodoscopic Shashlyk ECAL located at a distance $\gtrsim 10$ m from the target is shown for illustration; see, also,  e.g.,  
Ref. \cite{gams}. Note that (differently from the kaon case)  all $\pi^0, \eta$ and $\eta'$ mesons decay primarily in the target due to their extremely short lifetimes. The measured yield, corrected for the decay photon absorption in the target, directly 
gives the number  $n^{dec}_{\pi^0, \eta, \eta'}$ of $\pi^0, \eta$ and $\eta'$ decays in the target which   enter an equation analogous to Eq.(\ref{nev}):
\begin{eqnarray}
&n^{inv}_{\pi^0, \eta, \eta'}= k N_{\pi^-}Br(\pi^0, \eta, \eta' \to invisible)  \nonumber \\
& \simeq  \zeta \epsilon_{tag} Br(\pi^0, \eta, \eta' \to invisible) n^{dec}_{\pi^0, \eta, \eta'}.
\label{nevpi0}
\end{eqnarray}

\begin{table}
\caption{\label{tab:table2} Expected upper limits on the branching ratios of 
different decays into invisible final states calculated for the total number of $10^{12}$ incident 
pions or kaons( see text for details). }
\begin{ruledtabular}
\begin{tabular}{lr}
Expected limits on &  Present limit\\
the branching ratio & \\
\hline
Br$(\ksinv) \lesssim 10^{-8}$ & no~~~~~~~ \\
\hline
Br$(\klinv) \lesssim 10^{-6}$ & no~~~~~~~  \\
\hline
Br$(\pi^0 \to invisible )\lesssim 10^{-8}$ & $ < 2.7 \times10^{-7}$\cite{pi0}\\
\hline
Br$(\eta\to invisible )\lesssim 10^{-7}$ & $ < 1.0 \times10^{-4}$ \cite{bes14}\footnote{These limits are given in 
Ref.\cite{bes14} for the values $\frac{\Gamma(\eta (\eta') \to invisible)}{\Gamma(\eta (\eta') \to \gamma \gamma)}$
and  are re-calculated for  the ratios Br$(\eta(\eta') \to invisible )=\frac{\Gamma(\eta(\eta')\to invisible)}
{\Gamma(\eta(\eta')\to all)}$, respectively.}\\
\hline
Br$(\eta' \to invisible )\lesssim 10^{-6}$ & $ < 5.2 \times 10^{-4}$\cite{bes14}$^a$ \\
\hline
\hline
\end{tabular}
\end{ruledtabular}
\end{table}

The statistical limit on the sensitivity of the proposed experiment is mostly set  by the number of 
accumulated events. However, there is a limitation factor related to the  HCAL signal duration($\tau_{hcal} \simeq 100$ ns) which results in a  maximally allowed kaon counting rate $ \lesssim 1/ \tau_{HCAL} \simeq 10^{7} K^- /$s   above which a significant drop of signal efficiency due to the pileup effect is expected.
To evade this limitation, one could implement a special pileup-removal algorithm that allows for a 
high-efficiency  reconstruction of the  zero-energy signal properties and shape in high-pileup environments,  and then run the experiment at the rate $\simeq 1/\tau_{HCAL}\simeq 10^7~K^-/s$.
 Thus in the background-free experiment  one could potentially expect sensitivities in the $M^0 \to invisible$
 decay branching ratio that are even higher then those presented in Table II, assuming that 
the  exposure to the high-intensity kaon beam  is a few  months.

In the case of the $M^0 \to invisible$ signal observation,  several methods could be used to cross-check the result. For instance, 
to test whether the  signal is due to the HCAL nonhermeticity or not, one could perform  measurements with different HCAL thicknesses, i.e. with  one, two, three,  and four consecutive HCAL modules.  In this case the expected background level could be obtained by extrapolating the results to an infinite HCAL thickness.  
The evaluation of the signal and background  could also be  obtained from the results of measurements at different 
 beam energies. 

An interesting hypothetical question related to the test of the Bell-Steinberger relation is (if the signal is 
observed) whether it would it be possible in to test the CP violation in the invisible decays.  
It it clear that checking it directly (as in the case of the CP violating decay $K_L\to \pi \pi$) would be  difficult  because the final state is assumed to be unobservable. However, one can perform measurements to see if there is any variation 
of the zero-energy signal for  different lengths of the decay volume. For example, to cross-check whether the signal is mostly from the $K_S$ or $K_L$ decay, one could  remove the decay volume DV and put the HCAL calorimeter behind the veto system. This would not  affect the main background sources and  still allow for the production of $K_S$s , but  $K_L$ decays in front of the HCAL would be  suppressed. For measurements with large $L$ in order to ensure that there is no additional background due to the variation of the HCAL hermeticity, e.g. due to the large transverse 
fluctuation of the hadronic final state, or due to an unexpected (yet unknown)  $t$-dependence of the 
charge-exchange reactions at large $t$ the transverse HCAL size should be large enough. 
Finally, we note that the presented analysis gives an illustrative order of magnitude for the sensitivity of the 
proposed experiment and may be strengthened  by more detailed 
simulations of the  experimental setup.\\

\section {Conclusion}

Due to their specific properties, neutral kaons  are 
 still one of the most interesting probes of physics beyond the standard model  from both 
 theoretical and  experimental viewpoints. The Bell-Steinberger relation remains  
 the most sensitive probe of CPT invariance in the $K^0 - \overline{K}^0$ system. 
It connects CP and CPT violation in the mass matrix of the kaon system to CP and CPT violation in all decay channels of neutral kaons, assuming that there are no significant undiscovered decay modes of either $K_S$ or $K_L$, such as  decays into invisible final states.

In this work we   proposed performing an experiment dedicated to 
the sensitive search for the still unexplored invisible decays of neutral kaons, $\kslinv$ by using available 
30-100 GeV kaon beams from the CERN SPS. One of the goals of the proposed search is to clarify the  question of how much these  decays can influence the Bell-Steinberger analysis of the $K^0-\kob$ system.  
The experiment is also capable of searching  for $\pi^0, \eta, \eta' \to invisible$ decays with the SPS pion beams.
If  the $M^0 \to invisible$ decays  exist, they could be observed by looking for events with a 
unique signature: the total disappearance of the beam energy in afully  hermetic hadronic  calorimeter.
 A feasibility study of the experimental setup shows that this unique signal 
 of $M^0\to invisible$ decays allows for searches of  $\kslinv$ decays with a sensitivity in the branching ratio Br$(\kslbinv) \lesssim 10^{-8} (10^{-6})$, and 
$\pi^0,\eta, \eta' \to invisible$ decays with a sensitivity a few  orders of magnitude beyond the present experimental limits.
The sensitivitiues in the branching ratios Br$(\kslinv)$ are  significantly higher compared to the branching ratios of the  $K_S,K_L$ decay modes which contribute to the present accuracy of the Bell-Steinberger analysis  \cite{adpdg}. 

These results could be obtained with a detector that is  optimized for  
several of properties,  namely, i) the intensity and purity of the primary pion and kaon beams,  ii)  
high-efficiency  veto counters, and  iii) a high level 
of hermeticity in the hadronic  calorimeter are of importance.
Large amounts of high-energy hadrons and high background suppression are crucial to improving the sensitivity of the search. To obtain the best limits,  a compromise should be found between the background level and the energy and intensity  of the beam.  

The proposed experiment is complementary to the one recently proposed for a sensitive search for  dark photons 
 decaying invisibly to dark-sector particles at the CERN SPS \cite{sngldms, ldms}.
 It also  provides interesting  motivations for further kaon  studies and fits well with 
  the present  kaon physics program at CERN (see, e.g., Ref.\cite{ac}). 
\begin{center}
{\large \bf Acknowledgments}
\end{center}
I would like to  thank  A. Ceccucci, P. Crivelli,  N. Krasnikov, V. Matveev, V. Polyakov, and V. Samoylenko for useful discussions, and  A. Dermenev and M. Kirsanov  for their help in calculations.  The comments of G.D'Ambrosio and the encouragement of J. Steinberger are very much appreciated.

\end{document}